\documentclass[review]{elsarticle}

\usepackage{multirow}
\usepackage[table,xcdraw]{xcolor}
\usepackage{epsfig,amsmath,graphicx,amssymb,amsfonts,color,epsfig}

\biboptions{authoryear}
\usepackage{graphicx}
\usepackage[normalem]{ulem} 
\usepackage{soul}

\newtheorem{defn}{Definition}
\newtheorem{theorem}{Theorem}
\usepackage{hyperref}
\newcommand\remove[1]{}

\usepackage{subfig}
\journal{Journal of Mathematical Psychology}

\begin{document}

\begin{frontmatter}

\title{Preparation and Measurement in Quantum Memory Models}

\author{Mojtaba Aliakbarzadeh}

\address{ Information Systems School\\
       Queensland University of Technology\\
       Brisbane, 4000, Australia\\
       m.aliakbarzadeh@qut.edu.au}
       
\author{Kirsty Kitto}

\address{Faculty of Science and Engineering\\
       Queensland University of Technology\\
       Brisbane, 4000, Australia}
\address{Connected Intelligence Centre\\
		University of Technology Sydney\\
        PO Box 123, Broadway, 2007, Australia\\
        kirsty.kitto@uts.edu.au}

\begin{abstract}
Quantum Cognition has delivered a number of models for semantic memory, but to date these have tended to assume pure states and projective measurement. Here we relax these assumptions. A quantum inspired model of human word association experiments will be extended using a density matrix representation of human memory and a POVM based upon non-ideal measurements. Our formulation allows for a consideration of key terms like measurement and contextuality within a rigorous modern approach. This approach both provides new conceptual advances and suggests new experimental protocols.  
\end{abstract}

\begin{keyword}
Quantum cognition\sep Projective valued measurement (PVM)\sep Non-ideal measurement (POVM)\sep Density matrix\sep Semantic memory \sep Contextuality 
\end{keyword}

\end{frontmatter}

\newpage


\section{Introduction}

How should we model memory? As Shiffrin states: 
\begin{quote}
  \emph{ None of the models we
use in psychology or cognitive science, at least for any behavioral tasks I find to be of any
interest, are correct. We build models to increase our understanding of, and to slightly better approximate, the incredibly complex cognitive systems that determine behavior. } \citep{shiffrin:modeling}
\end{quote}
However, this pragmatism raises an interesting point. What do our models of memory assume? And how do they limit the way in which we can formulate a given memory model? 

Currently many models of Quantum Cognition (QC) apply a single state vector that assumes a system in a pure state \citep*{Aerts2007,Bruza2009,nelson.kitto.ea:how,PothosBusemeyer2013,BruzaKitto2015Probabilistic}. However, when we perform memory experiments we obtain ensemble data for a collection of subjects. This cannot be modeled with a pure state, rather a mixed state is required. In this work we will make use of the density matrix representation to model ensembles of human subjects in word association experiments. At first, we will provide a detailed technical description of Von-Neumann projective valued measurement (PVM). Although PVM measurement has been used in QC before, especially in the recall experiment of \citet{Bruza2009}, a better technical description is necessary to describe measurement on ensembles of subjects. Here we will make use of a more precise notation for the specific case of two observables in the recall experiment, and then we will generalize this notation for more possible senses. This will enable us to describe scenarios that have more possible outcomes for each observable. 

Another limitation of previous models for semantic memory in QC centers around the usage of projective measurement for cognitive systems. This is highly restrictive because QC (i) does not necessarily assume an orthogonal relationship between operators, and (ii) sometimes entails violations of repeatability. An analysis of these restrictions associated with the PVM formalism will lead us to introduce the more modern and general positive operator valued measure (POVM). We will show that this non-orthogonal measurement provides new understandings and extensions of the standard advantages of quantum inspired models of memory. 

Some existing research has already applied POVM in the construction of quantum models of cognition. For example, \citet*{KhrennikovDzhafarovBusemeyer} used POVM to model different arrangements of questions in opinion polling, including ``response (non)replicability'' and ``question order effect''. In another work, \citet{Khrennikov2014} employed POVM to describe a situation in which there are not sharp ``Yes/No'' answers to dichotomous decision observables. \citet{Yearsley2017AdvancedTool,YearsleyBusemeyer2016QuantumCognitionTutorial} provide a detailed tutorial for using POVM to model noisy and imperfect measurement, and their structure was used by \citet*{DenolfMartínez2017DilemmaGames} to model the prisoner’s dilemma experiment. During this period of emerging interest, we recognized that POVM could provide a natural model of the process of conceptual combination \citep{MojtabaKirsty2016}, and introduced a generalised Bell inequality, where POVMs were used to represent joint nonideal measurement for two observables. The current work will extend this early promising result, additionally describing a more general form of POVM for one observable.

We will show that the density matrix representation and POVM formulation suggest new sources of contextuality in the preparation and measurement processes respectively. This allows us to reconsider the interpretation of context within these new models. Although it is important to explain existing contextuality in cognitive experiments using better mathematical methods, we believe it is also essential to consider other sources for contextuality in those experiments; sources of contextuality that were ignored in previous work.

We also introduce another application of POVM in the modeling of memory. We will use Neumark’s dilation theorem to relate the full cognitive state of a subject to a restricted substate which represents only those cognitive processes through which a subject participates in an experiment. 

At the end of the paper, we will discuss a future direction that we believe will contribute to better understanding of cognitive states. We will point to a possible application in using Quantum Tomography to characterize the unknown state of a cognitive system. Using the insights that we gain from this characterization, we will suggest that a new experimental protocol could be created based on repeating projective measurements on similar ensembles of a subject to specify the unknown state of that subject. In an idealized situation, the whole parameters of an unknown cognitive state could be specified using a single POVM.

\section{The Quantum Model of Memory}\label{QuantumModelMemory}

Quantum Models of Memory \citep{Bruza2009,nelson.kitto.ea:how,BruzaKitto2015Probabilistic} treat words as states in a Hilbert space. The combined activation of words in memory is modeled using an entangled state, where an associative network is either fully activated, or not. In \citet{nelson.kitto.ea:how}, it was argued that associative semantic networks are constructed through the complex set of experiences that people undergo throughout their lifetime, and so are closely related to episodic memory, a point that opens up the possibility for linking episodic and semantic models of memory if we can construct more plausible relationships between them in our formalism. In particular, episodic memories beyond the boundaries of an experiment can be considered a form of experimental noise, a point that we will return to shortly.

Semantic associative models imply that the way in which a subject responds to a prime will affect their ability to recall other words not directly connected to that prime in e.g. a semantic network \citet*{Nelson2004}. This assumption can be tested experimentally,  and in \citet{BruzaKitto2015Probabilistic} a framework is provided for considering whether conceptual combination can be considered compositionally or not. Two tests are used to established compositionality; Marginal selectivity \citep{Dzhafarov2012} and a Bell type inequality. A number (21 from a total of 24) of conceptual combinations in that paper violated the marginal selectivity condition, while one of the combinations (BATTERY CHARGE) appears to satisfy marginal selectivity but violates a Bell-type inequality. More data is required before these results can be considered definitive. However, at this juncture it is a good idea to reconsider the theoretical apparatus of that model. Its reliance upon standard quantum models leaves it open to a number of criticisms from the perspective of psychology.
Indeed, for a number of reasons that will become apparent shortly we consider it important to extend that model to a more general and modern formulation. We will begin this extension with a move to the density matrix formulation.

\subsection{Constructing a density matrix}
\label{density}

We start with a consideration of the way in which a subject might recall an ambiguous word $A$ when cued with a particular prime. In quantum memory models this prime is represented as a basis state (i.e. a measurement context). Here we will use the eventualities $\{a'$, $a''\}$ to describe a subject's responses to a concept  $A$, which can be interpreted according to one of two possible dominant and subordinate \emph{senses}. When the dominant sense of concept $A$ is primed, and $A$ is interpreted in that sense by the human subject, then we designate $a' =+1$. If $A$ is not interpreted in that sense after priming the dominant sense, then we write $a' =-1$. Similarly, $a''=\{1, -1\}$ relates to situations where the subordinate sense of concept $A$ primed, and either recalled ($+1$) or not ($-1$).

  An example will help to make this formalism clear. Consider an experimental protocol where a subject cued with a concept $A$ (e.g. BOXER) using a word on a screen ``boxer''. According to the USF free association norms \citep{Nelson2004}, a subject is more likely to interpret BOXER in the sport sense than the animal sense. We term the sporting sense dominant and the animal sense subordinate. If a subject is first primed with the dominant sense of BOXER using the word ``glove", and then asked to interpret the concept BOXER, there is high possibility that they will recall a word that has a sport sense. This measurement process is represented by $\mathbb{A}'$, the result given by the subject is represented with $a'$, and $a' =+1$, as the response agrees with the way in which the subject was primed. If the subject interprets BOXER in another sense, then we write $a' =-1$. Conversely, if at first the subject is shown the word \textquotedblleft vampire\textquotedblright, then this is likely to awake the animal sense in the mind of human subject. When the subject responds in a way that agrees with the animal sense of the priming we write $a''=+1$, but if the concept is not interpreted in this subordinate sense, we use $a'' =-1$.  
 
Adopting von Neumann's approach to the quantum measurement of an idealised system using self-adjoint linear operators, we assume that an orthonormal basis exists. We can now construct a Hermitian matrix $\mathbb{A}$ as a series of projection operators \citep{Bruza2009}
  \begin{equation}{\label{vonneumann0}}
  	\begin{split}
  		\mathbb{A}=\sum_{k} a_{k}\mathbb{P}_{k}  
  	\end{split}
  \end{equation}  
where $\mathbb{P}_{k}$ is the projector onto the eigenspace of $\mathbb{A}$ with eigenvalue $a_{k}$, and each $a_{k}$ corresponds to the results of the measurement $\mathbb{A}$.
As an example, for two eigenvalues $a_{1}$ and $a_{-1}$, we can rewrite the Von-Neumann measurement as 
\begin{equation}{\label{vonneumannp1p-1}}
	\begin{split}
		{\mathbb{A}}=a_{1}\mathbb{P}_{1}+a_{-1}\mathbb{P}_{-1},
	\end{split}
\end{equation} 
where $\mathbb{P}_{1}$ and $\mathbb{P}_{-1}$ are the projectors onto the eigenspace of $\mathbb{A}$ for those two eigenvalues. 

For the case of the concept $A$ discussed above, we can therefore write out two noncommuting measurement operators $\{{\mathbb{A}}', {\mathbb{A}}''\}$ for the two different cases of priming (dominant and subordinate)
    \begin{align}{\label{A1A2operators}}    
  		{\mathbb{A}}'&=a_{1}'\mathbb{P}'_{1}+a_{-1}'\mathbb{P}'_{-1}
        \\
        &=a_{1}'|a_{1}'\rangle \langle a_{1}'|+a_{-1}'|a_{-1}'\rangle \langle a_{-1}'|,
        \\
  		{\mathbb{A}}''&=a_{1}''\mathbb{P}''_{1}+a_{-1}''\mathbb{P}''_{-1}
        \\
        &=a_{1}''|a_{1}''\rangle \langle a_{1}''|+a_{-1}''|a_{-1}''\rangle \langle a_{-1}''|.{\label{A1A2operatorsend}}     
  \end{align}
where $|a_{k}'\rangle$ and $|a_{k}''\rangle$ both represent potentialities of a subject's state of mind after priming. It is reasonable to consider these two operators $\{{\mathbb{A}}', {\mathbb{A}}''\}$ as noncommuting because we cannot prime with both dominant and subordinate senses simultaneously. This implies that the related projectors are noncommuting across the two primes.
  
  If we consider $|\psi\rangle$ as the cognitive state of a subject, then the probability of obtaining result $k$ is
  \begin{equation}{\label{pkprojective}}
  	\begin{split}
  		p(k)=\langle\psi|\mathbb{P}_{k}|\psi\rangle=Tr(\mathbb{P}_{k}|\psi\rangle\langle\psi|),
  	\end{split}
  \end{equation}
  and the subject's post-measurement state is
  \begin{equation}{\label{postprojective}}
  	\begin{split}
  		\frac{\mathbb{P}_{k}|\psi\rangle}{\sqrt{p(k)}}
  	\end{split}.
  \end{equation}
  This is known as a Projective Valued Measurement (PVM) in Quantum Mechanics (QM), a special class of general measurement which has the following properties \citep{Wheeler2012}:
  \begin{enumerate}[I.]
  	\item Hermitian: ${\mathbb{P}}^{\dagger}=({\mathbb{P}}^{T})^*$\\
  	A square matrix $\mathbb{P}$ is Hermitian if it is equal to its transposed complex conjugate.
  	This leads to an important property for operator $\mathbb{P}$, that its eigenvalues are real (not complex). 
  	\item Positive: $\langle\alpha|\mathbb{P}_{i}|\alpha\rangle\geq0$ (all $\alpha$)\\
  	Positivity allows us to treat the results of measurements as probabilities when coupled with the next property.
  	\item Complete: $\sum_{i}\mathbb{P}_{i}=\mathbb{I}$\\
  	The eigenvalues of a complete set of observables fully specify the state of a system. 
  	\item Orthogonal: $\mathbb{P}_{i} \mathbb{P}_{j}=\delta_{ij}\mathbb{P}_{i}$\\
   The results of measurement are completely independent from each other.
  \end{enumerate}  
  Returning to the scenario of priming the dominant sense, the probabilities of the measurement of  ${\mathbb{A}}'$ are given as follows:
  \begin{equation}{\label{A1probabilities}}
  	|\psi\rangle \xrightarrow[\text{dominant sense}]{\text{Prime with the}} \left\{
  	\begin{aligned}
  		&a'=+1 \quad \mbox{with probability} \quad |\langle a_{1}'|\psi\rangle|^2=\langle\psi|\mathbb{P}_{+1}|\psi\rangle\\\\
  		&a'=-1 \quad \mbox{with probability} \quad |\langle a_{-1}'|\psi\rangle|^2=\langle\psi|\mathbb{P}_{-1}|\psi\rangle
  	\end{aligned} \right.
  \end{equation}
Repeating this measurement multiple times allows us to calculate an expected value \citep{Wheeler2012}:
\begin{defn}
	The expected value of operator $\mathbb{A}$ acting on state $|\psi\rangle$ is calculated as  
	\begin{equation}{\label{means}}
		\begin{split}
			\langle \mathbb{A}\rangle_{\psi}=\sum_{k} a_{k} \langle\psi|\mathbb{P}_{k}|\psi\rangle= \langle\psi|\mathbb{A}|\psi\rangle, \quad where \quad k \in \{+1, -1\}
		\end{split}.
	\end{equation}
\end{defn}
We should note that any of the two measurement operators defined in (\ref{A1A2operators}--\ref{A1A2operatorsend}) can be placed instead of the operator $\mathbb{A}$ in this definition.

At this point we should ask whether such an experiment \emph{could} be completed multiple times. The state $|\psi\rangle$ denotes a cognitive state for a subject, and once we have performed the experiment we have irrevocably changed the state of our subject's mind to the state represented by \eqref{postprojective}. We will need an ensemble of subjects to repeat our experiment, but the proposition that even two subjects would share the same initial state $|\psi\rangle$ is highly unlikely. Although we might try to provide the same experimental conditions as we prepare our different subjects, we can not assume that they will all be in exactly the same state $|\psi\rangle$, this can happen because of unwanted priming effects or even the different dynamics of those subjects. Summing up all of our subjects, we can represent a scenario where some proportion of them are in the state $|\psi_1\rangle$, another proportion are in the state $|\psi_2\rangle$ and so on. Averaging these proportions with reference to our total subject pool would give us a scenario where 
\begin{equation}{\label{ensembleprobabilities}}
	\varepsilon(S(\psi_{1}, \psi_{2},...))\left\{
	\begin{array}{c}
		|\psi_{1}\rangle \quad \mbox{with probability} \quad p_{1}\\
		|\psi_{2}\rangle \quad \mbox{with probability} \quad p_{2}\\
		\vdots
	\end{array} \right.
\end{equation}
where we use $\varepsilon$ as an abbreviation for ensemble and $S$ for subject \citep{Wheeler2012}. 

For this expanded scenario, we can now rewrite the expected value for all of our measurements over this ensemble of subjects (using a standard approach that can be found in any QM text e.g. \cite{Wheeler2012}, or indeed in standard texts on QC e.g. \cite{Busemeyer2012}).
\begin{defn} The expected value for measurements over ensemble of subjects is calculated as	
	\begin{equation}{\label{densityensemble}}
		\begin{aligned}
			\langle \mathbb{A} \rangle_{\varepsilon}&=\sum_{v} p_{v} \langle \mathbb{A}\rangle_{\psi_{v}}\\
			&=\sum_{v} p_{v} \langle\psi_{v}|\mathbb{A}|\psi_{v}\rangle\\
			&=\sum_{j}\sum_{v} p_{v} \langle\psi_{v}|\mathbb{A}|e_{j}\rangle \langle e_{j}|\psi_{v}\rangle\\
			&=\sum_{j}\sum_{v} \langle e_{j}|\psi_{v}\rangle p_{v} \langle\psi_{v}|\mathbb{A}|e_{j}\rangle \\
			&=Tr({\rho}_{\epsilon}\mathbb{A}) \mbox{ where } \mathbb{\rho}_{\epsilon}=\sum_{v} |\psi_{v}\rangle p_{v}\langle\psi_{v}|.
		\end{aligned}
	\end{equation}
\end{defn}
While we started with orthogonal measurements, it is interesting that the state $|\psi_{v}\rangle$ which is used in the construction of the density matrix ${\rho}_{\epsilon}$ is not required to be orthogonal. As a result the different ensembles of states can lead to the same density matrix. The density matrix is mathematically equivalent to the vector representation of states, but provides a more convenient way to deal with some scenarios in QM, including the representation of ensembles of states \citep{Chuang}. The density matrix should have the following properties \citep{Wheeler2012}:
\begin{enumerate}[I.]
	\item Hermitian: ${\rho}^{\dagger}=(\rho ^{T})^*$ 
	\item Positive: $\langle\alpha|\rho|\alpha\rangle\geq0$ (all $\alpha$)
	\item Have unit trace: $Tr\rho=\sum_{v}p_{v}=1$
\end{enumerate}

The density matrix $\mathbb{\rho}_{\epsilon}$ in \eqref{densityensemble} becomes a \textit{pure} state $\mathbb{\rho}_{\psi}$, when one of the probabilities $p_{v}$ becomes equal to unity and the others vanish. This  signifies a return to the scenario where all subjects are prepared in the same initial state, i.e. we have
\begin{equation}{\label{purestate}}
	\begin{split}
		\mathbb{\rho}_{\psi}=|\psi\rangle\langle\psi|.
	\end{split}
\end{equation}
When we cannot make this simplifying assumption, we must consider $\mathbb{\rho}_{\epsilon}$ to be \emph{mixed}. In general, given a specific density matrix, we can discover if it is mixed or pure using the Trace, as $Tr(\rho^2)=1$ for pure states and $Tr(\rho^2)<1$ for mixed ones. When a system is in a pure state, both state vector and density matrix representations of a given system provide the same results \citep{Chuang}. 

Until now our description has been based on an idealized assumption that measurements are performed on the pure state $|\psi\rangle$ for each subject. In other words we have assumed that the cognitive state of a subject who recalls a concept $A$ can be represented by a pure state $|\psi\rangle$. But in reality we can not guarantee that the subject will adhere to our designed experimental protocol; it is clear that the human mind can process different concepts or events other than our intended concept $A$ during the experiment. The cognitive state of a human mind does not relate only to the process of recall. If we can represent the general state of the mind with a pure state $|\psi\rangle$, then the state that we use to model the recall process should be a subsystem of that pure state. In physics, as we mentioned earlier, we can use the density matrix to represent ensembles of states, however the density matrix can also be used to represent a subsystem of a pure state \citep{Chuang}. 

Elaborating, we will denote this extension of our formalism by rewriting the cognitive state of our subject as a composition of states labeled by $R$ and $E$, where $R$ denotes that part of the cognitive state that is directly influenced by the recall experiment and $E$ relates to the remainder. In physics, if a composite state can be represented as a pure state, then we cannot represent its subsystems using a pure state \citep{Chuang}. We can apply the reduced density matrix as
\begin{equation}{\label{reduceddensitymatrix}}
	\begin{split}
		\rho^R \equiv Tr_{E}(\rho^{RE}),	
	\end{split}
\end{equation}
where $\rho^{RE}$ describes the state of the composite system, and $Tr_{E}$ is an operator known as a partial trace on operator $E$. For example, to obtain the density matrix of subsystem $R$, we use the partial trace over subsystem $E$ \citep{Chuang}
\begin{equation}{\label{partialtrace}}
	\begin{split}
		Tr_{E}(|r_{1}\rangle\langle r_{2}|\otimes|e_{1}\rangle\langle e_{2}|) \equiv|r_{1}\rangle\langle r_{2}|Tr(|e_{1}\rangle\langle e_{2}|)
	\end{split}.
\end{equation}
where $\{|r_{1}\rangle,|r_{2}\rangle\}$ and $\{|e_{1}\rangle,|e_{2}\rangle\}$ are spanning vectors in the state space of $R$, and $ E$ respectively, and the standard trace operator  $Tr(|e_{1}\rangle\langle e_{2}|)=\langle e_{2}|e_{1}\rangle$ has been applied on the right hand side.

Now we can rewrite equations \eqref{pkprojective} and \eqref{postprojective} using the density matrix $\mathbb{\rho}$ of the ensemble of subjects or the subsystem $R$. The probability of obtaining the result $k$ becomes
\begin{equation}{\label{ensemblepkprojective}}
	\begin{split}
		p(k)=Tr(\rho\mathbb{P}_{k}),
	\end{split}
\end{equation}
and the state after measurement is
\begin{equation}{\label{ensemblepostprojective}}
	\begin{split}
		\frac{{\mathbb{P}_{k}}^{\dagger}\mathbb{\rho}\mathbb{P}_{k}}{p(k)}.
	\end{split}
\end{equation}
 
We note that the Hermitian matrix $\mathbb{A}$ that was defined in \eqref{vonneumann0} and used to construct the density matrix is not particularly comprehensive in its representation of memory, which makes it largely uninteresting in its current form. However, at this point it is possible to extend the basic approach, which considered only two possible senses. 

To this end, we note that it is frequently the case that  more than two interpretations are possible for one lexical  observable, a situation that can be represented by extending the set of projectors from our PVM measurement \eqref{vonneumannp1p-1} as 
\begin{equation}{\label{generalnonideal}}
	\mbox{Set of projectors for a PVM measurement}\left\{
	\begin{array}{l}
		\vdots\\
		\mathbb{P}_{k_{1}}\\
		\mathbb{P}_{k_{2}}\\
		\mathbb{P}_{k_{3}}\\
		\vdots
	\end{array} \right.
\end{equation}
For example, returning to the BOXER case discussed above, it is possible to interpret this ambiguous word in a third, clothing related, sense (e.g. BOXER SHORTS). This gives us three possible senses: ``sport'',``animal'' and ``clothing'' (but it is important to emphasize that this word could be interpreted in even more than these three senses). Similar to the original example, we can use the observable $\mathbb{A}'$ to represent a measurement process when the subject is first primed with the dominant sense and then  asked to interpret the word BOXER. In this case, $a'_{1}$ represents a case where the subject's response agrees with the primed sense, while $a'_{2}$ and $a'_{3}$ relate to two other possible responses in ``animal'' and ``clothing'' senses, and $a'_{4}$ represents all other possible senses. Then as in \eqref{A1A2operators} the Von-Neumann measurement for the observable $\mathbb{A}'$ becomes 
\begin{equation}{\label{vonneumannp1p2p3}}
	\begin{split}
		{\mathbb{A}'}=a'_{1}\mathbb{P}_{1}+a'_{2}\mathbb{P}_{2}+a'_{3}\mathbb{P}_{3}+a'_{4}\mathbb{P}_{4},
	\end{split}
\end{equation} 
where $\mathbb{P}_{1}$, $\mathbb{P}_{2}$, $\mathbb{P}_{3}$ and $\mathbb{P}_{4}$ are the projectors onto the eigenspace of $\mathbb{A'}$ for four eigenvalues $a'_{1}$, $a'_{2}$, $a'_{3}$ and $a'_{4}$. We will use this description of multiple senses to construct the generalized form of non-ideal measurement in the next section.

This section has presented a much more technical introduction to the model presented in \citet{Bruza2009}. We have shown that it is possible to generalize the standard quantum probability model using density matrices. This allows for the representation of scenarios where we cannot guarantee that an ensemble of subjects have all been prepared with the same pure cognitive state. This is an important consideration in psychology, the assumption that all subjects prepared in the same way are in the same pure cognitive state is a very strong one and does not match with reality. We have also introduced a second application of this density matrix apparatus to describe part of a larger cognitive system. The density matrix operator is capable of dealing with far more complexity in the quantum model of memory, that is, it  can fully characterize the state of a cognitive system. This means that once we are given the density matrix we can predict the outcome of any measurement on that state. But how we can access this knowledge about our system? In QM, the standard way to characterize the complete quantum state of a particle is by using quantum state tomography \citep*{wootters,Thew2002Qudit}. We will briefly describe how this method might be introduced to the field of QC in Section~\ref{tomography}, which identifies an unknown quantum state using a set of measurements. In the next section we will keep using the density matrix, where we will introduce a more general formalism for measurement in QC, demonstrating the specifics of how it can be used in quantum memory models.

\subsection{Non-ideal measurement (POVM)}
\label{POVM}

In reality, the process of recall does not always create sharp results like the von Neumann measurement described in the previous section assumes. This happens because of unwanted effects in the process of measurement. For example, there is no guarantee that a human subject will actually give a response corresponding to what they recalled. Despite the best experimental instructions, humans will be noisy in their responses. Thus, in the recall experiment described earlier for the word BOXER, a subject may be primed with the sport sense (the word ``glove'' was used in the previous example), think of ``Muhammad'', but censor their response giving a response with an animal sense instead (e.g.``dog'') with a probability $\xi$. This can be modelled using inefficient detectors. If we prepare the system in the state $\rvert +1\rangle$ (corresponding to the sport sense), then the result of measurement will be $\rvert +1\rangle$ with the probability $1-\xi$ and $\rvert -1\rangle$ with the probability $\xi$. To model this inefficiency, physicists often apply an unsharp measurement instead of an ideal von Neumann measurement \citep{Barnett2009,Wheeler2012,Muynck}.

Returning to the scenario where we prime the dominant sense (observable $\mathbb{A}'$), an ideal PVM measurement is described by the two projectors $\mathbb{P}_{+1}$ and $\mathbb{P}_{-1}$ introduced in \eqref{vonneumannp1p-1}. The possibility of imprecision arising in all subjects' recall processes is represented with the probability $\xi$. This means that if the result of measurement in the ideal situation was $+1$, a non-ideal situation would return $+1$ with the probability $1-\xi$ and $-1$ with the probability $\xi$. If the state of our system is described by the density matrix $\rho$, then the probability that the subjects recall words with the same sense as the prime is given by \citet{Barnett2009}
\begin{equation}{\label{probabilitiesof2results}}
	\begin{split}
		p(+1)=(1-\xi)Tr(\rho\mathbb{P}_{+1})+\xi Tr(\rho\mathbb{P}_{-1}).\\
	\end{split}
\end{equation}
By defining a new operator $\mathbb{\tilde{P}}_{+1}$ as 
\begin{equation}{\label{2pvms}}
\begin{split}
\mathbb{\tilde{P}}_{+1}=(1-\xi)\mathbb{P}_{+1}+\xi \mathbb{P}_{-1},\\
\end{split}
\end{equation}
we can use the technique in \citet{Barnett2009} to write the probability for the measurement outcome $+1$ in a manner similar to the PVM case \eqref{pkprojective}
\begin{equation}{\label{probabilitiesof2povm}}
	\begin{split}
		p(+1)=Tr(\rho\mathbb{\tilde{P}}_{+1}).\\
	\end{split}
\end{equation}
This $\mathbb{\tilde{P}}$ is a new type of operator. It is not required to be orthogonal and is used to describe this non-ideal situation. The operator is called a ``Positive Operator-Valued Measure'' or POVM.

It is possible to extend this situation, and to incorporate the multiple senses described earlier to reach the most general theoretical formulation using non-ideal measurement. As occurred for the case with two possible responses, we have to define probabilities for each non-ideal measurement. In this general case with $j$ possible responses, if the result of measurement in the ideal situation was $k$, a non-ideal situation would have the result $j$ with the probability $w_{j|k}$ where $\sum_{j} w_{j|k}=1$ for all $k$ \citep{Wheeler2012}. For a system in a state $\rho$, the probability of finding result $k$ after a non-ideal measurement would be represented by the following formula \citep{Wheeler2012}
\begin{equation}{\label{probabilitiesofnresultsPOVM}}
	\begin{aligned}
		p(k)&=\sum_{j} w_{j|k}Tr(\rho \mathbb{P}_{j})\\
		&=tr(\rho\mathbb{\tilde{P}}_{k}) \quad where \quad  \mathbb{\tilde{P}}_{k}=\sum_{j}w_{j|k}\mathbb{P}_{j}\\
	\end{aligned}
\end{equation}
The general form of equation \eqref{probabilitiesofnresultsPOVM} can be reduced to the simple form of equation \eqref{probabilitiesof2povm} if we have only two results, $+1$ and $-1$,  where $w_{j|k}$ can take two values $p$ and $1-p$. Thus it is possible to recover the simple scenario discussed above.  

Similar to the way in which the set $\{\mathbb{P}_{1}, \mathbb{P}_{2},...\}$ was a complete set of ideal measurements, the set $\{\mathbb{\tilde{P}}_{1}, \mathbb{\tilde{P}}_{2},...\}$ is a complete set of non-ideal measurements, but this time with the following properties \citep{Wheeler2012}
\begin{enumerate}[I.]
	\item Hermitian: ${\mathbb{\tilde{P}}}^{\dagger}=({\mathbb{\tilde{P}}}^{T})^*$
	\item Positive: $\langle\alpha|\mathbb{\tilde{P}}_{i}|\alpha\rangle\geq0$ (all $\alpha$)
	\item Complete: $\sum_{i}\mathbb{\tilde{P}}_{i}=\mathbb{I}$
	\item Typically non-projective and non-orthogonal: $\mathbb{\tilde{P}}_{i} \mathbb{\tilde{P}}_{j}\neq\delta_{ij}\mathbb{\tilde{P}}_{i}$
\end{enumerate} 
Any operators $\mathbb{\tilde{P}}_{1}, \mathbb{\tilde{P}}_{2},...$ that satisfy these properties are POVM.

The non-orthogonality of POVMs is a highly desirable feature in QC, especially for semantic memory models where there is no guarantee that the natural representation of multiple senses of a word should be orthogonal. Indeed, the senses of many words overlap to a degree often owing to shared background etymology.  This makes the assumption of orthogonal projective measurements in this class of model highly restrictive. 
Because of this desirable property of orthogonality, POVMs can bring new opportunities to model psychological phenomena that have not previously been modeled in PVM approaches. For example, it is possible that subjects give different responses to the same repeated cue, a scenario that could be termed  ``non-repeatability''. We have already modeled this property for recall experiment in an earlier work \citep{MojtabaKirsty2016}. Interestingly,  \citet{KhrennikovDzhafarovBusemeyer} also used POVM in opinion polling to demonstrate non-repeatability in an evolution-free framework.

To mathematically illustrate how non-repeatability arises within the POVM approach, we can rewrite $\mathbb{\tilde{P}}_{k}=\mathbb{A}_{k}^{\dagger}{\mathbb{A}_{k}}$
where $\mathbb{A}_{k}$ is called a measurement decomposition operator \citep{Jaeger2009}. In this case the state after measurement can be written
\begin{equation}{\label{postPOVM}}
	\begin{split}
		\frac{{\mathbb{A}_{k}}^{\dagger}\mathbb{\rho}\mathbb{A}_{k}}{\sqrt{p(k)}},
	\end{split}
\end{equation}
This post-measurement state indicates that unlike projective measurement, a repeated application of the POVM does not lead to the same result. We can quickly see this result if we apply the POVM observable once more on the post-measurement state of \eqref{postPOVM}, the result becomes
\begin{equation}{\label{postpostPOVM}}
	\begin{split}
		\frac{{\mathbb{A}_{k}}^{\dagger}{\mathbb{A}_{k}}^{\dagger}\mathbb{\rho}\mathbb{A}_{k}\mathbb{A}_{k}}{Tr({\mathbb{A}_{k}}^{\dagger}{\mathbb{A}_{k}}^{\dagger}\mathbb{\rho}\mathbb{A}_{k}\mathbb{A}_{k})}.
	\end{split}
\end{equation}
which is not necessarily equal to the previous state. It will be equal only if the POVM elements are idempotent (${\mathbb{\tilde{P}}}^2 = {\mathbb{\tilde{P}}}$). In this situation, POVM is reduced to a projective measurement. Despite this difference, there is a way to relate these two measurements to each other using Neumark's Theorem, which we will discuss in section \ref{Neumark}. 

Experimental scenarios can rapidly become very complex in the case of word association experiments. As we explained above, subjects may not report the word that sprang immediately to mind. A further complexity emerges where we consider the overarching social setting in which experimental priming is carried out \citep{MojtabaKirsty2016}. Even if we try our best to design an experiment with equally weighted primes for each possible sense, some primes are more dominant. For example, during an election period the ``political'' sense of the word PARTY may become stronger. A similar shift in weight towards the alternative sense might occur for a subject who went to a party the night before the experiment. We can apply this complexity to describe a non ideal choice of measurement settings in the generalized Bell-type experiments in cognition. In this case the priming with different senses occurs with different probabilities \citep{MojtabaKirsty2016}. 

The ideal Bell experiment modeled by \citet{BruzaKitto2015Probabilistic} assumes an equal choice of the different settings for any given subject (two operators $\{{\mathbb{A}}', {\mathbb{A}}''\}$ in (\ref{A1A2operators}--\ref{A1A2operatorsend}) which relate to the priming of two senses). It is possible to relax this assumption using the generalized Bell experiment. This is analogous to a situation where a biased interferometer leads a photon arriving at one of two detectors with different probabilities, which can be expressed by a bivariate POVM \citep{Muynck}. The model in that approach provides a joint non-ideal measurement of two observables, where for simplicity \citet{Muynck} assumes $100\%$ efficient detectors. This simplification removes the need to consider the possibility that the subject's response is not what first sprang to mind (i.e. the complexity of an inefficient detector). This allows us to assume ideal measurements for each observable separately and non-ideal measurement for the joint observable. We need the following definitions from \cite{Muynckpaper} to construct POVM as a \emph{jointly non-ideal measurement} of observables.
\begin{defn} {\label{Muynckeq1}} A POVM $\{\tilde{M}_m\}$ is a \emph{nonideal} measurement of the observable POVM $\{\tilde{N}_n\}$ if 
\begin{equation}
  \tilde{M}_m=\sum_n\lambda_{mn}\tilde{N}_n, \qquad \lambda_{mn}\geq 0, \qquad \sum_m \lambda_{mn} = 1.
\end{equation}
\end{defn}

\begin{defn} {\label{Muynckeq2}} Two observables ${\tilde{M}_{m}}$ and ${\tilde{N}_{n}}$ are simultaneously, or \emph{jointly measurable} if a bivariate POVM ${\tilde{R}_{mn}}$ exists such that its marginals $\{\sum_{n}\tilde{R}_{mn}\}$ and $\{\sum_{m}\tilde{R}_{mn}\}$ are POVMs and representing non ideal measurements of $\tilde{M}_{m}$ and $\tilde{N}_{n}$ respectively. 
\end{defn}

As was shown in \citet{MojtabaKirsty2016}, this approach can be applied to the experiment discussed by \citet{BruzaKitto2015Probabilistic}. Denoting the probability $\gamma$ for priming with the sense $\mathbb{A}'$, and the probability $1-\gamma$ for the priming with sense $\mathbb{A}''$, to represent the above scenario. As was the case in (\ref{A1A2operators}--\ref{A1A2operatorsend}), our observables $\{{\mathbb{A}}', {\mathbb{A}}''\}$ can be represented using two projectors ${\mathbb{P}_{1}, \mathbb{P}_{-1}}$. We can write the set of PVMs for the first and second observables as $(\mathbb{P}_{n}', \mathbb{P}_{m}'')$ where $n$ and $m$ take the values in $\{+1,-1\}$. The joint non ideal measurement for PVMs $(\mathbb{P}_{1}', \mathbb{P}_{-1}')$ and $(\mathbb{P}_{1}'', \mathbb{P}_{-1}'')$, can be constructed as a bivariate POVM \citep{DeMuynck2006,MojtabaKirsty2016}:  
\begin{equation}{\label{POVMmatrix}}
\begin{split}
\tilde{R}_{m n}^\gamma =
\begin{pmatrix}
0 & \gamma (\mathbb{P}_{1}') \\
(1-\gamma) (\mathbb{P}_{1}'') & \gamma (\mathbb{P}_{-1}') + (1-\gamma) (\mathbb{P}_{-1}''),
\end{pmatrix}
\end{split}
\end{equation}
The probability for this joint nonideal measurement is $p_{m n}=Tr\rho \tilde{R}_{m n}^\gamma$, as was the case for \eqref{pkprojective}. The top left hand corner of this matrix is equal to zero because the subject cannot be primed with two senses for a word at the same time. 

The marginals of $\tilde{R}_{m n}^\gamma$ are the POVMs $\{\tilde{M}_{m}\}=\{\gamma \mathbb{P}_{1}', I-\gamma \mathbb{P}_{1}'\}$ and $\{\tilde{N}_{n}\}=\{(1-\gamma)\mathbb{P}_{1}'',I-(1-\gamma)\mathbb{P}_{1}''\}$ which can be re+presented in matrix form as
\begin{equation}{\label{marginal1}}
\begin{split}
\begin{pmatrix}
\sum_{n}\tilde{R}_{1 n}^\gamma \\
\sum_{n}\tilde{R}_{-1 n}^\gamma
\end{pmatrix}=\begin{pmatrix}
\gamma & 0 \\
1-\gamma & 1
\end{pmatrix}
\begin{pmatrix}
\mathbb{P'}_{1} \\
\mathbb{P'}_{-1}
\end{pmatrix}
\end{split}
\end{equation}
\begin{equation}{\label{marginal2}}
\begin{split}
\begin{pmatrix}
\sum_{m}\tilde{R}_{m 1}^\gamma \\
\sum_{m}\tilde{R}_{m -1}^\gamma
\end{pmatrix}=\begin{pmatrix}
1-\gamma & 0 \\
\gamma & 1
\end{pmatrix}
\begin{pmatrix}
\mathbb{P''}_{1} \\
\mathbb{P''}_{-1}
\end{pmatrix}
\end{split}
\end{equation}
These marginals satisfy Definitions \ref{Muynckeq1} and \ref{Muynckeq2}. It is clear that the operators $\tilde{M}_{m}$ are $\tilde{N}_{n}$ are not commuting because $\mathbb{P}_{n}'$ and $\mathbb{P}_{m}''$ are not commuting. So we do not necessarily need commutativity of operators to construct non-ideal joint measurements \citep{DeMuynck2006}. Note that  $\tilde{R}_{m n}^\gamma$ only describes one concept (e.g. $A$ or $B$). This is unlike the scenario that arises for the standard Bell-type inequalities that were constructed using ideal joint measurements on commuting observables for both concepts $A$ and $B$ \citep{BruzaKitto2015Probabilistic}.

The direct product of the bivariate POVMs \eqref{POVMmatrix} for two concepts $A$ and $B$ in the Bell-type experiment of \citet{BruzaKitto2015Probabilistic} leads to a quadrivariate POVM, which can be written as
\begin{equation}{\label{POVMmatrix2}}
\begin{split}
\tilde{R}_{m_{A} n_{A} m_{B} n_{B}}^ {\gamma_{A}\gamma_{B}}= \tilde{R}_{m_{A} n_{A}}^ {\gamma_{A}} \tilde{R}_{m_{B} n_{B}}^ {\gamma_{B}}.
\end{split} 
\end{equation}
In this scenario there is no disturbing influence arising on the marginals of one concept when we change the measurement settings for another concept \citep{DeMuynck2006}. The measurement results of each concept are influenced by the measurement settings of that concept (complementarity). This enables the POVM formalism to model contextual behavior without making use of nonlocality. Complementarity then provides us with a local explanation for violations of the generalized Bell inequality which is expressed using the quadrivariate probability distribution \citep{DeMuynck2006}
\begin{equation}{\label{probabilityquadrivariate}}
\begin{split}
p_{m_{A} n_{A} m_{B} n_{B}}^ {\gamma_{A B}}= Tr \rho \tilde{R}_{m_{A} n_{A}}^ {\gamma_{A}} \tilde{R}_{m_{B} n_{B}}^ {\gamma_{B}}.
\end{split}
\end{equation}
Recall in this generalized form of the Bell experiment, each subject comes to the experiment with a different historical context. This context affects the way in which subjects are primed, as the semantic network will activate differently in response to the prime \citep{nelson.kitto.ea:how}. This activation relates to the probability $\gamma$ that we used to construct the POVM in \eqref{POVMmatrix}. A violation of the generalized Bell inequality would occur because of each subject's unique historical context.
 
  In this section we have described two methods for constructing a POVM, the first for one observable and the second for two. While we have shown that it is possible to construct different POVMs for these specific experimental scenarios, note that the properties of POVMs mentioned earlier imply that each POVM $\mathbb{\tilde{P}}$ is unique to the relevant experimental context. Much more work remains to catalog other psychologically plausible mechanisms that can arise in quantum memory experiments, and to demonstrate how they might be modelled using an approach based upon POVM. 
 
\section{Advantages of using a density matrix and POVM approach}\label{Advantages}

At this point we have described the two processes of preparation and measurement for a quantum memory model. In QM they are considered as separate processes. For example, as \citet[p.154--p.155]{Isham2001} states:
\begin{quote}
  \emph{A measurement is an operation on a system that probes that quantum state immediately before the measurement is made... state preparation is an operation whose aim is to force the system to be in some specified state immediately after the operation.} 
\end{quote}
In this paper we introduced the density matrix as a practical tool for describing preparation when dealing with an ensemble of subjects (in Section \ref{density}). This operator $\rho$ most generally represents a mixed state and it contains all the information necessary to predict any possible measurement outcome. For the measurement process we introduced a POVM which gives us a more realistic depiction of word association experiments than an approach based on standard projective measurement (as defined in Section \ref{POVM}). \citet{Franco2016Towards} recently constructed a quantum inspired model of decision making which follows a similar methodology; treating  preparation as a process where information is provided to a subject, and  the measurement stage as a process of testing subjects at the end of this preparation phase. 

Preparation and measurement have other more specific applications. For example, as we described in the previous section, Muynck's joint non-ideal measurement can be used to describe the situation of complementarity as it arises in the generalized Bell inequality. We will now introduce some more specific applications, explaining how they can serve to advance the field of QC.

\subsection{Neumark's theorem}\label{Neumark}

To draw attention to the different roles of PVM and POVM measurements in a cognitive experiment we can use Neumark's Theorem \citep{PeresAsher1990}. This theorem provides a tool for dealing with noise in a cognitive experiment (of the type that was described in Section \ref{QuantumModelMemory}). In that experiment we considered the state of a subject's mind as the composition of two states ``R'' and ``E'', where ``R'' denotes that part of the cognitive state related to recall experiment, and ``E'' is considered as the remainder, which we will refer to as noise. Noise arises from events or thoughts outside the defined bounds of the experiment (e.g. what the subject ate for breakfast, or an accident they witnessed on the way to the experiment). 

Neumark's theorem relates the POVM of state ``R'' to a projective measurement on the composition of states ``R'' and ``E'' \citep{Muynck}. In other words, it extends the Hilbert space of ``R'' to the tensor product space of ``R'' and ``E''.
\begin{theorem}[Neumark]
	An arbitrary POVM on a Hilbert space $H_{R}$ can be expressed using a PVM in a larger Hilbert space H containing $H_{R}$.
\end{theorem}
The inverse situation arises in a similar manner to the partial trace that we introduced in \eqref{reduceddensitymatrix}: Given any PVM on a Hilbert space $H$, we can find a POVM on a subspace $H_{R}$. In fact, we create the POVM on a sub-system when we do not need to consider the extra information contained within the higher dimensional Hilbert space. Tracing out this noise using the mathematical structure of POVM requires that we understand cognitive function well enough to construct the appropriate PVM and POVM for the Hilbert space $H$ and subspace $H_{R}$ respectively. Thus, we would need to understand the variables involved in shaping states ``R'' and ``E''. Current experimental developments may not provide us with this ability; as \citet{Khrennikov2014} discuss, the brain can be both a system and the observer in a QC system, which can make it difficult to isolate from its mental environment.

Progressing in this area will require significantly more work to understand how boundaries should be defined in cognitive experiments. For example, we are immediately confronted with the question of: what should be considered noise in a given experiment? 

It is essential that the field of QC  consider the effects of noise in our models for different cognitive experiments. Our work in Section \ref{POVM} was just a first step in this direction. We have so far considered two cognitively motivated effects of noise as: (1) the probability $\xi$, which represents the imprecision across all subjects in a recall experiment; and (2) the probability $\gamma$ for priming with the sense $\mathbb{A}'$ and $\mathbb{A}$ in the generalized Bell experiment. However, more work remains to be completed before it will be possible to construct comprehensive models for memory using modern quantum inspired methods.

\subsection{An operational approach to modelling cognition-defining context}
\label{Operationalapproach}

Much of this section was inspired by the work of \citet{spekkens2005contextuality}, who has generalized the standard treatment of contextuality in quantum theory to arbitrary operational theories. We have used his approach to treat contextuality as it occurs for both preparation and measurement processes in cognition. This approach extends existing works on contextuality in QC by \citet*{Aerts2013}; \citet{BruzaKitto2015Probabilistic,Bruza2016Syntax}; \citet*{DzhafarovKujalaLarsson2015,DzhafarovKujalaContext-Content2016}. Thus, we have provided: (a) a method for modelling contextuality in preparation, and; (b) refinements in our understanding of contextuality as it arises during measurement. 

Our approach here has been deliberately formulated in terms of basic operations such as preparation $(P)$ and measurement $(M)$, and the probabilities for various possible measurement outcomes. In physics, operational theory is defined based on these experimental procedures \citep{HarriganSpekkens2010Einstein,AbramskyHeunen2013,spekkens2005contextuality}. This section will apply the same approach to cognitive experiments. More specifically, we will show how this approach is compatible with the process of recall demonstrated earlier (in Section \ref{density}), and allude to the manner in which the same approach can be used to model contextuality. 

The mathematical structures we need for the operational approach were introduced earlier in Sections \ref{density} and \ref{POVM}. To define this approach, we apply the density matrix $\rho$ to model the preparation process $P$, and a POVM$\{\mathbb{\tilde{P}}_{k}\}$ to the measurement process $M$ as follows \citep{AbramskyHeunen2013,spekkens2005contextuality}: 

\begin{defn} {\label{Operationldef}} Operational theory defines the probabilities $p(k|P,M)$ of different outcomes $k$ given specific preparation and measurement procedures.
\end{defn} 
To fully describe the Spekkens approach we require a precise understanding of what he means by an Ontological model. Here, the intrinsic properties of a physical system are called its \emph{ontic state} and is denoted by $\lambda$ (where $\lambda$ belongs to a set of all possible ontic states $\Lambda$). To use Spekkens' approach in cognition, we will make use of a similar conception: an ontic  state should refer to the reality of the cognitive system, that is, the presumed features of a cognitive state (of mind) which exist without performing experiments or any other form of observation. (Our definition of ontic state in cognition will be illustrated further on  with reference to the recall experiment.)

Here we bring the definition of an ontological model of operational theory as suggested by \citet{spekkens2005contextuality} and \citet{HarriganSpekkens2010Einstein} to define a notion of contextuality:

\begin{defn} 
{\label{ontologicaldef}} An ontological model (of operational quantum theory) posits an ontic state space $\Lambda$ and for every preparation procedure $P$ over $\Lambda$ appoints a probability distribution $\mu_{P}(\lambda)$. Similarly, a probability distribution $\xi_{M,k}(\lambda)$ is attributed over the different outcomes $k$ of a measurement $M$ for every ontic state $\lambda \in \Lambda$. Finally, to be consistent with operational theory for all $P$ and $M$, an ontological model should satisfy
\begin{equation}{\label{OperationalOntological}}
\begin{split}
p(k|P,M)=\sum_{\lambda \in \Lambda} \xi_{M,k}(\lambda)\mu_{P}(\lambda)
\end{split}
\end{equation}
\end{defn}

This definition makes it possible to specify the notion of a contextual model, where instead of explicitly considering quantum states and POVMs, we specify probabilistic  interpretations of the preparation and measurement procedures that are used to create them ($\mu_{P}(\lambda)$ and $\xi_{M,k}(\lambda)$). These are the  probabilities that determine what can be known and inferred by observers  \citep{HarriganSpekkens2010Einstein,spekkens2005contextuality}.

Having provided the mathematical details of an operational theory and ontological model, now we can study the Spekkens approach to contextuality  \citep{spekkens2005contextuality}. The fundamental idea of noncontextuality in this approach is that processes which are operationally equivalent, should not be distinguishable in an ontological model \citep{Leifer2014}. This  means two processes which generate the same observable probabilities, should be represented by the same probability distributions over their underlying ontic state. Thus, we can say the two preparation processes of a state are noncontextual when they yield the same probability distributions without changing the intrinsic properties of the system. 
 
 We can illustrate this approach in cognition with reference to the recall experiment that was introduced earlier in Section \ref{density}. In that example, the preparation scenario leads to some proportion of subjects belonging to different cognitive states \eqref{ensembleprobabilities}, where the density matrix of ensembles of subjects is represented using the convex composition of pure states $\psi_{v}$ representing those subjects, with the probability $p_{v}$.\footnote{It worth pointing out that we can also consider a more general scenario that the density matrix of this recall experiment as a convex composition of mixed states.}

This density state of ensembles is related\footnote{It is possible to have a one-to-one relationship between the state of a system $\psi$ and the reality, although in most of physical models $\psi$ only indicates a state of incomplete knowledge about reality. A more complete discussion can be found in \citet{HarriganSpekkens2010Einstein}.} to the preparation process. However, we know that our cognitive reality is not completely controlled by the preparation process, which means that preparing our cognitive system in a specific state $\rho$ does not give us any information about the exact ontic state. Our knowledge of the ontic state can only be described by the probability distribution  $\mu(\lambda)$ that we introduced earlier. The assumption of noncontextual preparation entails that the probability distribution of preparation $P$ on an ontic state $\lambda$ depends only on the $\rho$ related to $P$ \citep{spekkens2005contextuality}
\begin{equation}{\label{ontologicalequivalentpreparation}}
	\begin{split}
		\mu_{P}(\lambda)=\mu_{\rho}(\lambda).
	\end{split}
\end{equation}
This implies that for an ensemble of subjects to be noncontextual, the distribution $\mu_{P}(\lambda)$ should not depend on a specific convex decomposition of $\rho$ (each convex decomposition provides a different context for the preparation $P$). In other words, consider a hypothetical scenario where two preparation procedures result in the same density matrix. This density matrix of the ensemble of subjects can be implemented based on the two\footnote{It is clear in a more general scenario, there could be more possible convex decompositions for the density matrix $\rho$.} different convex decompositions of $\rho$, represented as $\rho=\sum_{v}p_{v}\rho_{v}$ (see equation \eqref{densityensemble}). Each of these possible decompositions can be associated to a member of the equivalence class
of preparation procedures for density matrix $\rho$. These two procedures are noncontextual if the probability distribution $\mu_{P}(\lambda)$ is independent of each member of that class.

As with the preparation process, we can consider the measurement process as measuring the ontic state of our cognitive system. As for preparation, performing this measurement does not guarantee access to the ontic state, it only provides  different probabilities that the system exists in one of a collection of ontic states. These probabilities are represented by $\xi(\lambda)$ \citep{spekkens2005contextuality}. The assumption of noncontextual measurement entails that the probability distribution of measurement $M$ on an ontic state $\lambda$ depends only on the POVM $\{\mathbb{\tilde{P}}_{k}\}$ related to $M$:
\begin{equation}{\label{ontologicalequivalentmeasurement}}
	\begin{split}
		\xi_{M,k}(\lambda)=\xi_{\{{\tilde{P}}_{k}\},k}(\lambda).
	\end{split}
\end{equation}

 One way to represent context for measurement in cognition is to consider Neumark's theorem (Section \ref{Neumark}). This approach suggests that each way of recognizing a POVM on subsystem $H_{R}$, based on its coupling with the environment (noise) and performing a projective measurement on a composite system $H$, implicitly generates a context. In other words, different sources of noise can lead to different contexts. Examples of these sources of noise might include episodic memories constructed for different subjects throughout their lifetimes (and represented in a semantic network), e.g. subjects who have had their breakfast or not, give different response to a priming word ``food''. The measurements of psychology are inherently noisy, but this approach offers ways in which we might start to model this phenomenon.

In this section, we have shown that reconsidering an ensemble of subjects and a generalized measurement for cognitive systems leads us to identify new sources of contextuality. This approach describes the way in which contextuality affects the preparation process for the first time in cognition. Moreover, the suggested contextuality model for measurement process extended the standard measurement approaches traditionally used in QC with a  non projective measurement. Furthermore, this approach opens up a possibility for extending this model using contextuality models in physics, where every convex decomposition of a POVM $\{\mathbb{\tilde{P}}_{k}\}$ reveals a context for the measurement process \citep{spekkens2005contextuality}.

\section{What is a quantum cognitive state?}
\label{Discussion}

In this paper we first generalized the process of preparation and measurement for cognitive systems, and  then discussed some possible advantages of this representation. However, it is important to realize that much of the mathematics utilized in QC rests upon rather shaky foundations. To further improve quantum inspired models of cognition we need to advance in our mathematical understanding of the most basic cognitive states. As an example, the model provided in Section~\ref{density}, of the two senses that a subject might associate with the ambiguous word BOXER is constructed using the assumption of a basic two-level cognitive state. The subject can give many possible answers to an observable $\mathbb{A}$ (in this case the cue word BOXER), but we assume in our models that they fall into one of two possible senses (i.e. the same as the priming sense and different from the priming sense), which are denoted with +1 and -1. This basic state can be considered similar to a single qubit (a spin-1/2 particle) system in QM. It is represented by a pure state
\begin{equation}{\label{aqubitdirac}}
\begin{split}
|\psi\rangle=\alpha|0\rangle +\beta|1\rangle
\end{split} 
\end{equation}
where $|\alpha|^2$ and $|\beta|^2$ are probabilities of spins up and down respectively, and the state of the qubit is a vector in a two-dimensional complex vector space with the orthonormal computational basis $|0\rangle$ and $|1\rangle$. One of the main differences between a qubit and a classical bit is this superposition property. Unlike a classical bit, which acts akin to a coin with only two possible states of ``heads'' and ``tails'', a qubit can exist in any weighted continuum of states between $|0\rangle$ and $|1\rangle$. However, when it is measured a qubit still only gives outcomes indicative of $|0\rangle$ or $|1\rangle$ \citep{Chuang}. This property has been widely exploited in QC \citep*{Busemeyer2012,ZhengBusemeye2013ThePotential,Aerts2007,MasanariKhrennikov2015Adaptivity}. For example, \citet{PothosBusemeyer2013} represent the happiness of a hypothetical person using the superposition state $|\psi\rangle=a|happy\rangle+b|unhappy\rangle$. After being asked about her happiness, and the subject deciding upon her answer, the state vector becomes $|\psi\rangle=|happy\rangle$ or  $|\psi\rangle=|unhappy\rangle$.

However, an approach like this leaves us with few ideas as to what this representation of $|\psi\rangle$ actually symbolises. What is the underlying cognitive state? And how does it evolve in time as a person moves through their day? Here, we will provide some guidelines that could be considered in future research aimed at clarifying the representation of cognitive states.  We note that many more questions are provided in this section than answers, but consider it appropriate to draw attention to what is an underexplored but important area for future research.

\begin{figure}[htp]
    \centering
    \includegraphics[width=0.25\textwidth]{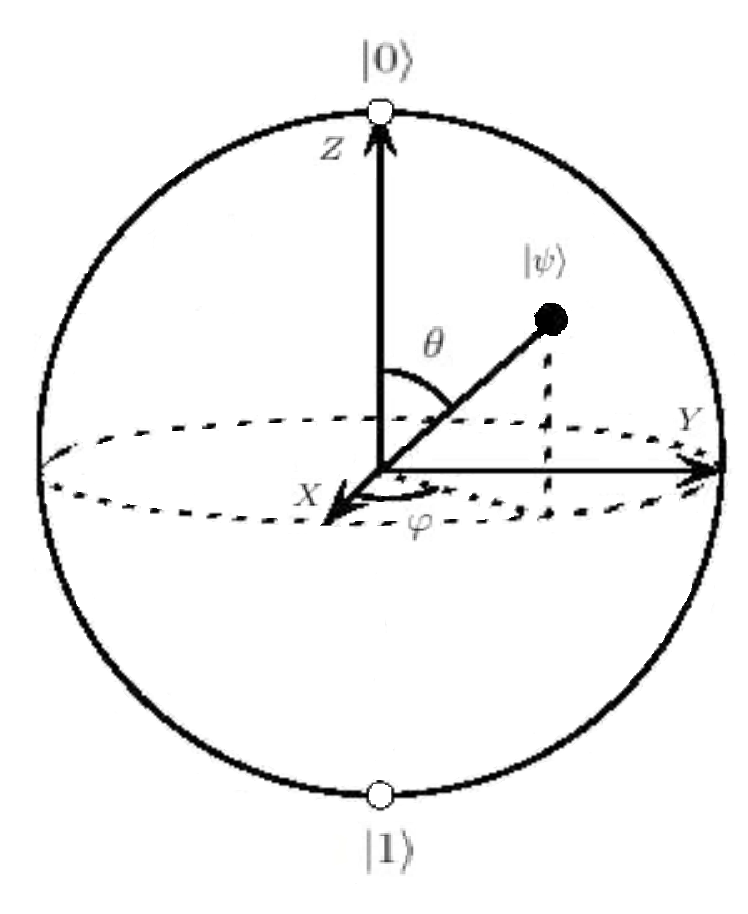}
    \caption{Bloch sphere representation of a qubit.}
    \label{fig:Bloch}
\end{figure}

Rather than representing qubits using the abstraction of a complex vector space, it is possible to more fully visualize their properties using the geometrical Bloch sphere representation. This method also provides a more explicit representation of the types of operations that we can apply on a qubit. We carry out this transformation by rewriting equation \eqref{aqubitdirac} as
\begin{equation}{\label{aqubitgeometric}}
\begin{split}
|\psi\rangle=\cos\frac{\theta}{2}|0\rangle +\sin\frac{\theta}{2}e^{i\varphi}|1\rangle,
\end{split} 
\end{equation}
where spherical coordinates are defined by latitude $\theta$ and longitude $\varphi$. Each pure state represented by equation \eqref{aqubitdirac} associates to a point on the surface of a unit sphere in the Euclidean $3$-dimensional space (see figure \ref{fig:Bloch}). 
In this geometrical representation, two orthogonal basis states $|0\rangle = |\uparrow\rangle$  and $|1\rangle = |\downarrow\rangle$ correspond to an orientation of the spin in $+z$ and $-z$ directions respectively. And superposition states can correspond to other orientations of the spin in different spatial directions, e.g. the state $|0_{y}\rangle$ is oriented in the $+y$ direction. The Spherical polar coordinates $(r,\theta,\varphi)$ can be related to Cartesian coordinates $(x,y,z)$ by
	\begin{equation}{\label{SpjericalCartesian}}
		\begin{aligned}
			x&=r\sin \theta \cos\varphi\\
			y&=r\sin \theta \sin\varphi\\
			z&=r\cos \theta.
		\end{aligned}
	\end{equation}
To measure spin of the qubit in each of these directions we would apply the Pauli matrices
\begin{equation}{\label{Pauli}}
	\begin{split}
\sigma_{x}=\begin{pmatrix} 
0 & 1 \\
1 & 0 
\end{pmatrix}, \quad
\sigma_{y}=\begin{pmatrix} 
0 & -i \\
i & 0 
\end{pmatrix}, \quad
\sigma_{z}=\begin{pmatrix} 
1 & 0 \\
0 & -1 
\end{pmatrix}
\end{split}.
\end{equation}
In other words, we can visualize the two-level quantum system using this three-dimensional Bloch sphere representation. A cognitive state built from this basic representation brings with it the possibility of modelling its dynamical evolution, but still has a direct mapping to the qubits which have been used in the previous work in QC. However, the visualization of more than two-level quantum systems can have different geometries and usually need higher-dimensional Bloch based vector representation \citep*{Kimura2003TheBloch,BertlmannKrammer2008Bloch,GoyalSimon2016Geometry} \footnote{It is not always necessary to consider a higher dimension, as an example, \citet*{KurzyńskiKołodziejski2016Three-dimensional} suggest a three-dimensional visualization for qutrit.}. Likewise, cognitive systems of more than two-levels  (like the example we described in equation~\eqref{vonneumannp1p2p3}) could still be modelled by these higher-dimensional representations.

However, to employ this Bloch sphere representation of a qubit in cognition, we have to provide appropriate meanings for these three directions in phase space. \citet{YearsleyPothos2013Challenging} provide an example interpretation in a decision making experiment, by projecting a bivariate observable on $Z$ direction. This work employs a Bloch sphere representation to build up an understanding of how a cognitive system might evolve in time to define a test for violations of the temporal-Bell inequalities. Similarly, \citet*{BroekaertBasieva2017Quantum-likeDynamics} provide the geometric interpretation for the evolution of states implied by Hamiltonian. Their Hamiltonian is built based on only two Pauli matrices to describe different dynamical evolution scenarios. However, in both of these approaches, the derivation of the Hamiltonian of the system makes reference only to the mathematical aspects of the Bloch sphere, without providing  an exact cognitive meaning for the directions $x$, $y$ and $z$ that are used in these models.

This is an important point to emphasize; there is little connection in these more advanced models between the physical formalism and the cognitive meaning. This paper has been careful to associate a strong psychological interpretation to the more advanced models of measurement and preparation as they apply to cognition. However, to move forwards we will require a far stronger connection to the underlying meaning of a cognitive qubit, which we will term a \emph{cobit}. As we emphasized earlier, it is necessary to provide cognitive meanings for the different Cartesian directions in the geometrical representations used with this approach. However, to transform this representation back to the complex vector space, we should still be able to provide cognitive meanings for the orthonormal computational basis $|0\rangle$ and $|1\rangle$ (like the two senses that we assigned to these two basis states in equation \eqref{aqubitdirac}). We admit that this geometrical interpretation does not scale to more complex multipartite situations in a straightforward manner, which potentially limits its utility as a general model, however, we consider it necessary that QC place more of an emphasis upon finding the underlying dynamical representation of cognitive processes, and a model based upon a high dimensional Bloch based representation is an immediately plausible option, as has already been recognised by a number of papers previously published in this domain. Providing the cognitive meaning for a cobit as discussed above, helps us to interpret the cognitive meanings of applied operators on that cobit. As an example, we could reach a better understanding of the rotations group $SO(3)$ and $SU(2)$ and their special relation with each other \footnote{The special orthogonal group $SO(3)$ represents rotations around the origin of a three-dimensional Euclidean space and the special unitary group $SU(2)$ is the set of 2 by 2 unitary matrices with determinant 1 (like the set of Pauli matrices described in equation \eqref{Pauli}). The relation between these two rotations groups is an one-to-two correspondence between any $R \in SO(3)$ and $2U \in SU(2)$ \citep{Miller1972SymmetryGroups}.}. While an unsatisfactory lack of detail still remains in mapping such models to cognitively plausible representations, our mapping in this paper of the density matrix to interpretable semantic memory tasks gives a new avenue that will help  to link this more interpretationally robust class of models to the extensive data sets that have been collected in this the domain of memory and recall already (see e.g.  \citet{nelson.kitto.ea:how} for a summary of one such dataset). We leave this contribution for future work.

If we can successfully create a cognitively well justified geometrical representation for our cobit then it will be straightforward to extend the model for multiple cobits. As \citet{Chuang} explain, treating qubits as abstract mathematical objects enables us to generalize the concept for more complex situations (e.g. multiple qubits) without depending upon a specific realization. This geometrical representation opens different avenues to make use of the approaches identified in this paper for using the density matrix and generalized measurement in modelling semantics. As an example of a further approach that this avenue might open up, we will briefly introduce the concept of cognitive tomography, and suggest a way in which it might be used to characterize unknown cognitive states.
 
\subsection{Cognitive Tomography}
\label{tomography}

 In a QC model, the result of measurement gives an indication of the state of a subject's mind with reference to a measurement scenario, or question, just before the measurement occurred. For instance, in the BOXER example described in Section \ref{density}, each answer $+1$ or $-1$ indicates whether BOXER would be interpreted in the same way as the priming occurred (as represented by the operators $\mathbb{A}'$ or $\mathbb{A}''$), or not.                                                                                                                                                       
But to fully understand the state of a subject's mind when faced with the word BOXER, we cannot rely on one measurement alone. One possibility would be to repeat the measurement many times on the same subject to get an average of different results. A memory experiment would need to repeat the cuing procedure in a variety of different contexts. However, this demonstrates precisely how difficult it is to construct a reliable measurement in cognition, because the response a subject gives to an experiment can affect the response that they give for the following experiments (see Section \ref{density}). 

So how can we proceed in finding a more precise understanding of the underlying cognitive state? One possibility would be to apply a method inspired by Quantum Tomography, which specifies an unknown quantum state by performing measurements over an ensemble of equally prepared identical quantum states \citep{Leonhardt}. To use this method for cognition, we would need to provide the same experimental conditions as we prepare our different subjects \footnote{This does not necessarily lead to an ensemble of subjects with the same pure cognitive state, as we mentioned in Section \eqref{density}.}. This is  similar to the scenario in physics where a device produces a beam of spin-1/2 particles \citep{wootters}.  It is possible to predict the spin state of particles that the device produces if we perform a set of orthogonal measurements in the $x, y$ and $z$ directions. These three measurements should be ``mutually conjugate'' \citep{wootters} which means that an eigenvector of any one of them must be an equal superposition of the eigenvectors of the two others. For this set of measurements, each different measurement provides information independent from the information provided by the other measurements \citep{wootters}.

A detailed mathematical description of tomography for cobits would rely upon having a precise cognitive meaning for the $x, y$ and $z$ directions in their geometrical representation. Then to estimate a cognitive state of cobits we would need to repeatedly apply three linearly independent observables (associated with those three directions) on three sub-ensembles of subjects \citep*{Wootters1987,wootters,gibbons2004discrete}. These observables would need  to be ``informationally complete", and so completely specify the state of the system. Thus, this method holds promise for being able to help us to estimate the unknown state of a cognitive state. An advantage of this approach is that it allows us to express a cognitive state as a real function on a discrete phase space instead of using the common method of the density matrix. This real function which is known as a Wigner function behaves as a probability distribution, but it can take negative values \citep{wootters,gibbons2004discrete}. There are recent interests of using negative probabilities in cognition to model a decision making experiment  \citep{BarrosOas2014NegativeProbabilitiesCounter} or even to model contextuality \citep*{DeBarrosKujala2015}. As an alternative to these approaches we could consider the Wigner function \citep*{RaussendorfBrowne2017ContextualityWigner,DelfosseOkay2017EquivalenceContextualityWigner,KenfackYczkowski2004NegativityWignerNonclassicality} which has been widely used in
nonclassical calculations in QM. This method could potentially be extended to more complex situations of multiple cobits tomography, with an associated increase in the number of necessary observables. To reduce the number of measurements required to specify the description of multiple cobits, we could use POVM instead of the projective measurements \citep*{Lundeen,wootters}. There is also a possibility of extending this method to the more ambitious scenarios of ``cotrit'' (three-level cognitive system) or ``codit''  ($d$-level cognitive system) tomography based on qutrit and qudit tomography in physics \citep{Thew2002Qudit}. Thus, the examples like that modelled in \eqref{vonneumannp1p2p3} could be more completely specified using this approach, an avenue that we leave to future work.

\section{Conclusion}

In this study we assumed that the measurement process for a cognitive system is separate from its preparation process. We provided a detailed mathematical description of these two processes by introducing density matrices and non-ideal measurement. Having created this more rigorous approach we applied it to existing concepts in QC such as complementary and contextuality, as well as investigating how it might be extended to new concepts like tomography.

We believe these approaches provide a better quantum inspired models of cognition, and so could lead to a better understanding of cognitive systems. As an example, in Section \ref{Operationalapproach}, the model of contextuality based on Spekkens' operational method provides a new way to study this important phenomenon in cognition. This model is more comprehensive than previous studies because of its consideration of the preparation process and non-ideal measurements. 

This work provides us with a number of new avenues to follow as we attempt to develop a more detailed understanding of the complex process of cognition, specifically, memory and recall. It suggests a way in which we could start to approach the problem of modelling an underlying cognitive state, and so work towards plausible models of the various ways in which these evolve as a person interacts with the world \citep{nelson.kitto.ea:how}. The array of episodic events that each of us takes part in every day all influence our underlying cognitive state, and it is essential that we develop modelling methodologies that are capable of capturing the full complexity of this important process. Adopting an operational approach to QC offers precisely this opportunity.

\end{document}